\documentclass[runningheads]{llncs}
\usepackage[T1]{fontenc}
\usepackage{cite}
\usepackage{url}
\usepackage{amsmath,amssymb,amsfonts}
\usepackage{graphicx}
\usepackage{textcomp}
\usepackage{xcolor}
\usepackage{diagbox}
\usepackage{tablefootnote}
\usepackage{bm}
\usepackage{bbding}
\usepackage{algorithmic}
\usepackage{algorithm}
\usepackage{multirow}
\usepackage{multicol}
\usepackage{enumitem}
\usepackage{booktabs}
\usepackage{subfigure}
\usepackage{hyperref}
\usepackage[misc]{ifsym}
\begin{document}
\title{ParamSpMM: Adaptive and Efficient Sparse Matrix-Matrix Multiplication on GPUs for GNNs\thanks{This manuscript is an author-posted version of a paper accepted by the 9th International Workshop on Graph Data Management and Analysis (GDMA 2025), held in conjunction with the 30th International Conference on Database Systems for Advanced Applications (DASFAA 2025).}}
\titlerunning{ParamSpMM}
%
\author{Lixing Zhang \and
    Guanhua Ye \and
    Hongzheng Li \and
    Shigang Li \and
    Yingxia Shao \Letter
}
\authorrunning{Lixing et al.}
%
\institute{Beijing University of Posts and Telecommunications, Beijing, China \\
    \email{\{zhanglixing, g.ye, Ethan\_Lee, lishigang, shaoyx \}@bupt.edu.cn}}

\maketitle              
\begin{abstract}
    Fueled by the ability to mine real-world graph data, GNN applications have experienced phenomenal growth. Sparse Matrix-Matrix Multiplication (SpMM) is a critical operator in GNNs. However, existing SpMM designs for GNNs struggle to adapt to diverse input characteristics. In this paper, we first conduct a comprehensive analysis of existing SpMM optimizations, revealing their limitations through statistical and empirical evidence. Based on this analysis, we introduce ParamSpMM, a parametric approach for highly adaptive and efficient SpMM computation in GNNs. It incorporates a new data structure, the Parameterized Compressed Sparse Row (PCSR), to flexibly integrate existing optimization techniques. ParamSpMM enables the configuration of these optimization techniques according to various input characteristics. Furthermore, we complement ParamSpMM with an ML-based SpMM-decider that predicts optimal configurations based on carefully crafted input features. Our evaluations demonstrate that ParamSpMM outperforms Nvidia cuSPARSE with an average speedup of $\mathbf{1.92\times}$, significantly enhancing GNN training efficiency.

\end{abstract}
\section{Introduction}
Graph Neural Networks (GNNs) have emerged as a vital tool in graph data mining, showing advantages in traffic prediction \cite{GNN:traffic:guo2019attention}, recommendation systems \cite{GNN:recommendation}, chemistry \cite{GNN:chemistry}, and biomedicine \cite{GNN:biology}. GNNs are renowned for their scatter-and-gather paradigm, where each node aggregates the neighbors' embeddings and updates its own embedding accordingly. This computation can be modeled as Sparse Matrix-Matrix Multiplication (SpMM), where the sparse matrix is the graph's adjacency matrix, and the dense input matrix corresponds to the embedding matrix.

SpMM often accounts for the majority of GNNs' execution time, making it a primary bottleneck that requires optimization \cite{huang2020ge, GNN_survey}. However, the diversity of input graphs and embedding dimensions complicates SpMM optimization for GNNs in three aspects:

\textbf{Various data locality of graphs.} If the neighboring nodes in a graph have close ID numbers, the graph will benefit from better data locality, resulting in a higher cache hit ratio in sparse linear computation~\cite{arai2016rabbit,wang2021gnnadvisor}. However, the presence of diverse community structures in real-world graphs\cite{fortunato2010community-detect} leads to various data locality.

\textbf{Various degree distributions of graphs.} There are Poisson degree distributions found in road traffic networks \cite{chandrasekaran2009survey:traffic_network} and Power-law degree distributions observed in social networks \cite{sala2010brief:power-law}. Graphs with the latter distribution suffer more from workload imbalance in SpMM.

\textbf{Various embedding dimensions.} GNNs have a wide range of network architectures with varying dimensions of node embeddings across layers. SpMM in GNNs thus involves dense input matrices with diverse column dimensions.

However, existing SpMM designs lack the flexibility to address the above three aspects of diversity, offering limited customization of the SpMM kernel for specific inputs. Static kernels in popular vendor-provided frameworks like DGL~\cite{DGL} and PyG~\cite{PYG} cannot adapt to different inputs. On the other hand, existing dynamic designs~\cite{wang2021gnnadvisor, fan2023fast, huang2020ge, ASpT:hong2019adaptive, dai2022heuristic-DASpMM} use fixed optimization strategies, providing limited adaptability by tuning a few parameters focusing on limited aspects of the whole performance issues.

The limitations of existing dynamic designs are evident in two aspects: \textbf{(1) Incomplete exploitation of optimization opportunities.} Some works~\cite{ASpT:hong2019adaptive} employ blocking to utilize data locality but ignore workload imbalance caused by skewed degree distributions. Others~\cite{fan2023fast, huang2021understanding, wang2021gnnadvisor, dai2022heuristic-DASpMM} focus on workload balancing without adequately utilizing data locality. \textbf{(2) Poor adaptability to diverse SpMM inputs.} Existing works struggle with unexpected inputs, demonstrating a mismatch between applied optimization techniques and the input characteristics. The blocking technique in ASpT \cite{ASpT:hong2019adaptive} becomes inefficient on ultra-sparse matrices due to excessive zero padding. Some work \cite{wang2021gnnadvisor, fan2023fast} apply workload balancing by default, degrading performance on graphs with relatively balanced degree distribution.

In this paper, we first provide an in-depth analysis of existing SpMM optimizations and reveal their limitations in handling diverse SpMM inputs. Based on the analysis, we introduce ParamSpMM, a parametric approach for SpMM computation in GNNs. The basic idea under the hood of ParamSpMM is to integrate the three existing optimization strategies—blocking, workload balancing, and thread coarsening—to enable high adaptability across diverse SpMM inputs, as each optimization strategy is tailored to a specific aspect of SpMM input diversity. We design an adaptive representation for the sparse matrix, called Parameterized Compressed Sparse Row (PCSR) format, to enable the seamless integration of the three optimizations. Based on the new format, ParamSpMM provides adaptive and customizable high-performance SpMM kernels via flexible parameter configuration, that responds to the various input characteristics.

While ParamSpMM offers comprehensive configuration options, manually identifying the optimal configuration remains challenging. To fully capitalize on the characteristics of diverse SpMM inputs, we introduce an ML-based SpMM-decider for ParamSpMM to predict the optimal configurations. With carefully crafted features across size, degree distribution, and data locality, SpMM-decider automates complicated configuration tasks, thereby augmenting ParamSpMM with self-driven capability.

Extensive evaluations demonstrate that ParamSpMM secures an $\mathbf{1.92\times}$ average speedup over Nvidia cuSPARSE \cite{naumov2010cusparse}. Embedded into GNNs training, ParamSpMM demonstrates an average of $\mathbf{1.96\times}$ speedup over DGL \cite{DGL} across various GNN models and datasets. We outline contributions as follows:
\begin{itemize}[leftmargin=*]
    \item We conduct an in-depth investigation into the limitations of existing SpMM optimization methods, revealing their inefficiencies in handling diverse data localities, degree distributions, and node embedding dimensions.
    \item We develop ParamSpMM, a flexible parametric framework for SpMM computation that dynamically adjusts to the varying characteristics of SpMM inputs in GNNs. This approach enables high-performance SpMM kernel customization, offering significant adaptability and efficiency across diverse input scenarios.
    \item We propose an ML-based SpMM-decider for the optimal configuration of ParamSpMM to efficiently leverage the SpMM input characteristics.
\end{itemize}

\section{Background}\label{background}
In this section, we introduce the basic Compressed Sparse Row (CSR) based SpMM. Notably, SpMM is defined as $\bm{A}_{n\times n} \bm{B}_{n\times dim}=\bm{C}_{n\times dim}$, involving the sparse matrix $\bm{A}$ and dense matrices $\bm{B}$ and $\bm{C}$. $dim$ is $\bm{B}$'s column dimension.

\textbf{CSR format.} CSR represents a sparse matrix via three arrays—$rowPtr$, $colIdx$, and $val$—facilitating efficient traversal of nonzero elements within a sparse row. The traversal range of $i$-th row is specified by $(rowPtr[i],rowPtr[i+1])$. Based on CSR, a prevalent approach for computing SpMM on GPUs is in a row-wise manner, where a worker (e.g., a thread warp of $\omega$ threads) computes a segment of $\bm{C}[i][:]$ while traversing nonzeros in $\bm{A}[i][:]$.

\textbf{Thread Mapping.} A commonly employed thread mapping scheme arranges thread blocks into a 2D grid~\cite{yang2018design_principles}. A thread block consists of $\mathcal{W}$ thread warps, each identified by $(blk.x, blk.y, warpId)$. Warps with the same $(blk.x,warpId)$ but different $blk.y$ compute different $\omega$-length segments in the same $\bm{C}$ row.

\textbf{CSR-based SpMM.} The essence of SpMM computation is the multiply-accumulate (MAC) operation, which multiplies a nonzero from $\bm{A}$ with a corresponding element in $\bm{B}$ and accumulates the result to $\bm{C}$. Given that $dim$ is a multiple of $\omega$, Algorithm \ref{Alg:CSR_SpMM} describes the computing process from a thread's perspective. The algorithm first determines the output position in $\bm{C}$ (Lines 1-5). Then, it establishes the traversal range (Lines 6-7), which also indicates the current warp's workload. For each iteration, a MAC operation is performed by retrieving $\bm{B}$'s data based on the $colIdx$ of the traversed nonzero (Lines 9-11). Finally, the results of the accumulated segments are written to $\bm{C}$ (Line 13).

\begin{algorithm}[t]
    \renewcommand{\algorithmicrequire}{\textbf{Input:}}
    \renewcommand{\algorithmicensure}{\textbf{Output:}}
    \caption{Basic CSR-based SpMM on GPUs}
    \label{Alg:CSR_SpMM}
    \begin{algorithmic} [1]
        \REQUIRE{$rowPtr$, $colIdx$, $val$, $\mathcal{W}$}
        \ENSURE{$\bm{C}$}
        \STATE {$warpId = threadId / \omega$}
        \STATE {$laneId = threadId \% \omega$}
        \STATE \textcolor{blue}{/*position of a computed segment in $\bm{C}$*/}
        \STATE {$Crow = blk.x * \mathcal{W} + warpId$}
        \STATE {$seg=blk.y*\omega+laneId$}
        \STATE {$head = rowPtr[Crow]$}
        \STATE {$tail = rowPtr[Crow+1]$}
        \FOR{ $i$ in $[head, tail)$}
        \STATE \textcolor{blue}{/*one multiply-accumulate (MAC) job*/}
        \STATE {$Brow, v = colIdx[i], val[i]$}
        \STATE{$res+=v*\bm{B}[Brow][seg]$}
        \ENDFOR
        \STATE{$\bm{C}[Crow][seg + i*\omega]=res$}
    \end{algorithmic}
\end{algorithm}

\vspace{-10pt}
\section{Analysis of Existing SpMM Optimizations}
In this section, we provide relevant statistics and empirical findings that highlight the limitations of existing optimizations. These observations underpin our design of ParamSpMM. Details on the experimental setup are provided in Section \ref{exp_setup}.

\vspace{-10pt}
\subsection{Analysis of Blocking Technique}\label{opp_blocking}
Basic CSR-based SpMM has irregular memory access patterns for matrix $\bm{B}$ due to the non-uniform layout of nonzeros in $\bm{A}$ as demonstrated in Lines 10-11 of Algorithm \ref{Alg:CSR_SpMM}. To mitigate this issue, blocking techniques are employed in sparse kernels \cite{li2022efficient-magicube,ASpT:hong2019adaptive,yesil2022-DDB,vldb-SpMV} to reduce irregular memory accesses.

Nonzeros in the same column of $\bm{A}$ exhibit identical memory access patterns to $\bm{B}$ during MAC operations. By grouping these nonzeros for processing by a single thread, we can effectively exploit $\bm{B}$'s data reuse through registers or shared memory. For graph datasets with extreme sparsity ($\mathbf{99.99\%}$)~\cite{hu2020OGB,leskovec2014snap,bader201110thDIMACS}, vectorized blocking (e.g., $2\times 1$) is more efficient than 2D blocking (e.g., $2\times 2$) by minimizing zero padding and thus reducing unnecessary computations. Thus, we adopt vectorized blocking in this work, organizing $\bm{A}$'s nonzeros into vectors based on a configurable Vector Size ($\mathcal{V}$).

\textbf{Observation: The effectiveness of vectorized blocking varies with graph data locality.} Inappropriate $\mathcal{V}$ choices can lead to excessive zero padding and performance degradation. Table \ref{tab_VS} shows SpMM throughput with different $\mathcal{V}$ on four representative real-world graphs when $dim=32$, with the corresponding zero padding ratios listed in parentheses. For $\mathcal{V}=1$, the padding ratio is $\mathbf{0\%}$, as it introduces no zero padding. Notably, with $\mathcal{V}=3$, $\mathbf{97.5\%}$ of our 202 graphs show padding ratios above $\mathbf{50\%}$, indicating over half of the computation is redundant. Therefore, in this paper, the domain of $\mathcal{V}$ is limited to $\{1,2\}$. The optimal $\mathcal{V}$ and padding ratios vary by graph, highlighting the necessity of adaptive $\mathcal{V}$ for blocking tailored to each graph’s data locality.

\begin{table}[h]
    \centering
    \vspace{-10pt}
    \caption{Throughput in GFLOPS under Various $\mathcal{V}$, with zero padding ratios listed in parentheses.}
    \vspace{-5pt}
    \label{tab_VS}
    \resizebox{0.9\columnwidth}{!}{%
        \begin{tabular}{c|cccc}
            \toprule
            \textbf{graphs}          & \textbf{coPapersCiteseer} & \textbf{coPapersDBLP}  & \textbf{sx-askubuntu} & \textbf{sx-mathoverflow} \\ \midrule
            \textbf{$\mathcal{V}=1$} & 1794  (0\%)               & 1672 (0\%)             & \textbf{386} (0\%)    & \textbf{865} (0\%)       \\
            \textbf{$\mathcal{V}=2$} & \textbf{2170} (26.8\%)    & \textbf{1870} (30.6\%) & 337 (47.8\%)          & 767 (49.0\%)             \\
            \textbf{$\mathcal{V}=3$} & 1635 (39.6\%)             & 1470 (44.6\%)          & 318 (64.3\%)          & 700 (65.6\%)             \\
            \bottomrule
        \end{tabular}%
    }
    \vspace{-15pt}
\end{table}

\subsection{Analysis of Workload Balancing Technique}\label{opp_balancing}
\vspace{-5pt}
The workload assigned to a thread warp is defined as $l=tail-head$, representing the traversal range of nonzeros (Lines 5-6 of Algorithm \ref{Alg:CSR_SpMM}). Workload imbalance in SpMM stems from uneven $l$ distribution across sparse rows.

The nonzero-split approach for workload balancing redistributes workloads across warps to mitigate imbalances in sparse computations~\cite{yang2018design_principles, merrill2016mergeSpMV, wang2021gnnadvisor, fan2023fast, dai2022heuristic-DASpMM}. It redistributes the workload of a heavily loaded thread warp across multiple warps by splitting nonzeros within a sparse row. While this improves workload balance, it compromises memory efficiency as each warp must independently write its partial results to the same output segment, resulting in repeated writes of some results. In contrast, SpMM without balancing writes each result to $\bm{C}$ only once.

\begin{figure}[b]
    \centering
    \includegraphics[width=1.0\linewidth]{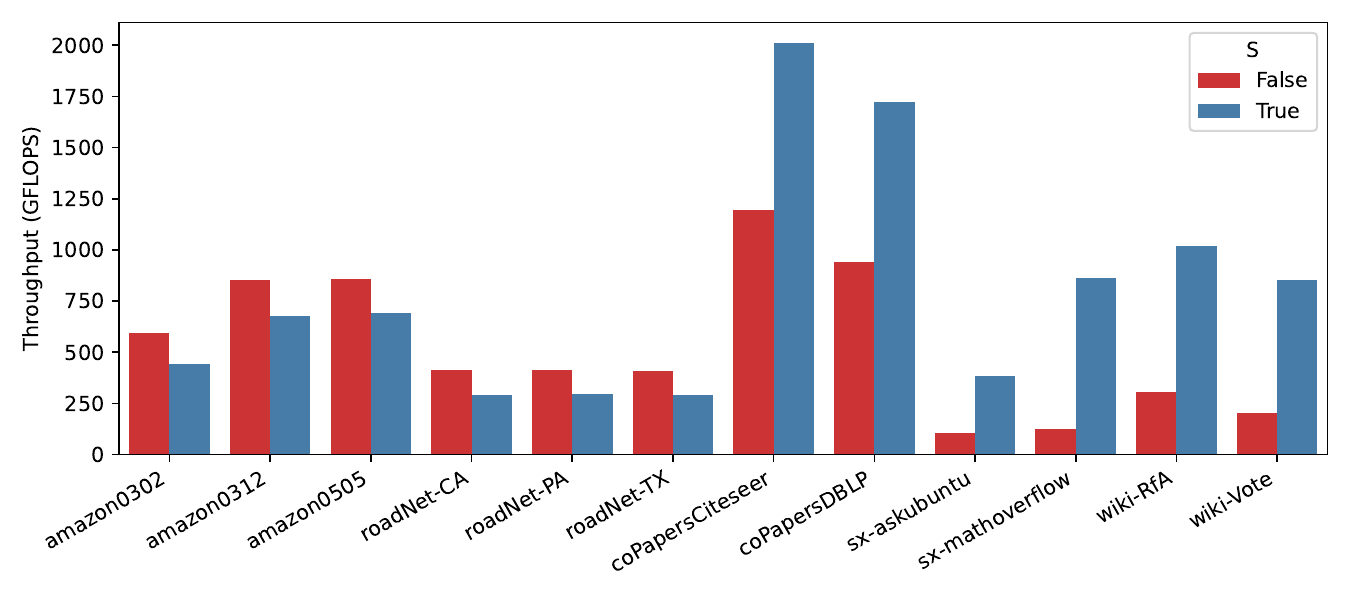}
    \vspace{-30pt}
    \caption{Throughputs of SpMM with or without workload balancing.}
    \label{fig_SP}
    \vspace{-10pt}
\end{figure}

\textbf{Observation: The effectiveness of workload balancing varies with graph degree distribution.} Applying balancing on graphs with relatively balanced degree distributions may not yield performance improvements, owing to increased memory writes and additional overhead from balancing-related bookkeeping instructions. Figure~\ref{fig_SP} demonstrates the throughputs (GFLOPS) of SpMM with and without workload balancing on 12 real-world graphs when $dim=32$. For simplicity, we use $\mathcal{S}$, a boolean value, to represent whether to apply balancing. The effectiveness of balancing varies across graphs, necessitating adaptive balancing. This observation highlights the importance of the adaptive application of workload balancing based on graph degree distribution.

\vspace{-10pt}
\subsection{Analysis of Thread Coarsening Technique}\label{opp_CF}
Thread coarsening~\cite{yang2018design_principles, huang2020ge, volkov2008benchmarking-coarsening, wang2021gnnadvisor, fan2023fast} allocates a set of independent MAC jobs to a single thread for each nonzero traversal iteration, and the size of the set is determined by Coarsening Factor ($\mathcal{F}$). Each warp reuses the fetched value and column index of a nonzero from $\bm{A}$ in one iteration to execute $\mathcal{F}*\omega$ MAC jobs, producing a length of $\mathcal{F}*\omega$ segment results in $\bm{C}$, with $\mathcal{F} \in  [1,CEIL(\frac{dim}{\omega})]$.

Thread coarsening brings a theoretical $\mathcal{F}$-fold reduction in the memory transactions for $\bm{A}$'s data and improves instruction-level parallelism (ILP). However, it increases register usage and reduces active threads on the fly~\cite{huang2020ge}. Therefore, an improper $\mathcal{F}$ potentially degrades performance.

Additionally, when $dim$ is not a multiple of $\omega*\mathcal{F}$, the residual warp (computing the last segment of a $\bm{C}$ row) may have a MAC-job gap compared to other warps with the same $(blk.x, warpId)$ but different $blk.y$. This MAC-job gap occurs under an improper $\mathcal{F}$ and wastes computing resources, which is overlooked by existing works. We quantify it as $gap_{\mathcal{F}}$:
\begin{align}
    \vspace{-20pt}
    tn^{\mathcal{F}}_{dim}  & = \min(dim, \mathcal{F} \cdot \omega) \notag                     \\
    tr^{\mathcal{F}}_{dim}  & = dim \mod (\mathcal{F} \cdot \omega) \notag                     \\
    gap^{\mathcal{F}}_{dim} & = tn^{\mathcal{F}}_{dim} - tr^{\mathcal{F}}_{dim} \label{eq:gap}
    \vspace{-20pt}
\end{align}
where $tn^{\mathcal{F}}_{dim}$ is the ideal MAC-job number under chosen $\mathcal{F}$ and $dim$, and $tr^{\mathcal{F}}_{dim}$ is the actual MAC-job number for the residual warp.

\textbf{Observation: The effectiveness of thread coarsening varies with $\bm{dim}$ and graphs.} Given $dim\in \{64,96,128,160\}$, Table \ref{tab:DCP} presents the distribution of optimal $\mathcal{F}$ settings across 202 real-world graphs. N/A means that the value of $\mathcal{F}$ is out of range and is thus exempted for a given $dim$. The corresponding MAC-job gap for each $dim$ is also presented in parentheses. The results reveal: (1) Improper $\mathcal{F}$ degrades performance due to MAC-job gaps for a given $dim$. $\mathcal{F}=2$ for $dim=96$, $\mathcal{F}=3$ for $dim=128$, and $\mathcal{F}=2,3,4$ for $dim=160$ are less preferred due to their wider MAC-job gaps compared to other $\mathcal{F}$ options. (2) Despite some $\mathcal{F}$ demonstrating the same MAC-job gap (e.g., $\mathcal{F}=1,2$ when $dim=64$), identifying the optimal $\mathcal{F}$ for a specific sparse matrix (graph) is non-trivial, involving memory transactions, local resource usage, and parallelism. These findings necessitate adaptive $\mathcal{F}$ selection based on both $dim$ and graph characteristics.

\begin{table}[t]
    \small
    \centering
    \caption{The Distribution of Optimal $\mathcal{F}$ Setting}
    \label{tab:DCP}
    \resizebox{0.6\columnwidth}{!}{%
        \begin{tabular}{c|c|c|c|c}
            \toprule
            \diagbox{$\mathcal{F}$}{$dim$} & \textbf{64} & \textbf{96}  & \textbf{128} & \textbf{160}  \\ \midrule
            1                              & 20.30\% (0) & 24.26\% (0)  & 21.29\% (0)  & 25.74\% (0)   \\
            2                              & 79.70\% (0) & 1.98\%  (32) & 36.14\%  (0) & 5.94\%  (32)  \\
            3                              & N/A         & 73.76\% (0)  & 0\%   (64)   & 6.44\%   (32) \\
            4                              & N/A         & N/A          & 42.57\%  (0) & 0\%  (96)     \\
            5                              & N/A         & N/A          & N/A          & 61.88\% (0)   \\
            \bottomrule
        \end{tabular}
    }
    \vspace{-15pt}
\end{table}

\vspace{-10pt}
\subsection{Summary}
As analyzed above, existing optimizations inadequately accommodate diverse SpMM inputs. A comprehensive design that flexibly integrates these optimizations is crucial for constructing a robust SpMM kernel accommodating three aspects of input diversity. Therefore, we introduce ParamSpMM with key parameters: $\mathcal{V}$ for vectorized blocking, $\mathcal{S}$ for workload balancing, $\mathcal{F}$ for thread coarsening, and $\mathcal{W}$ to determine the thread block size. These parameters collectively enable fine-tuned optimization, adapting to various input characteristics.

\begin{figure*}[b]
    \centering
    \includegraphics[width=1.0\linewidth]{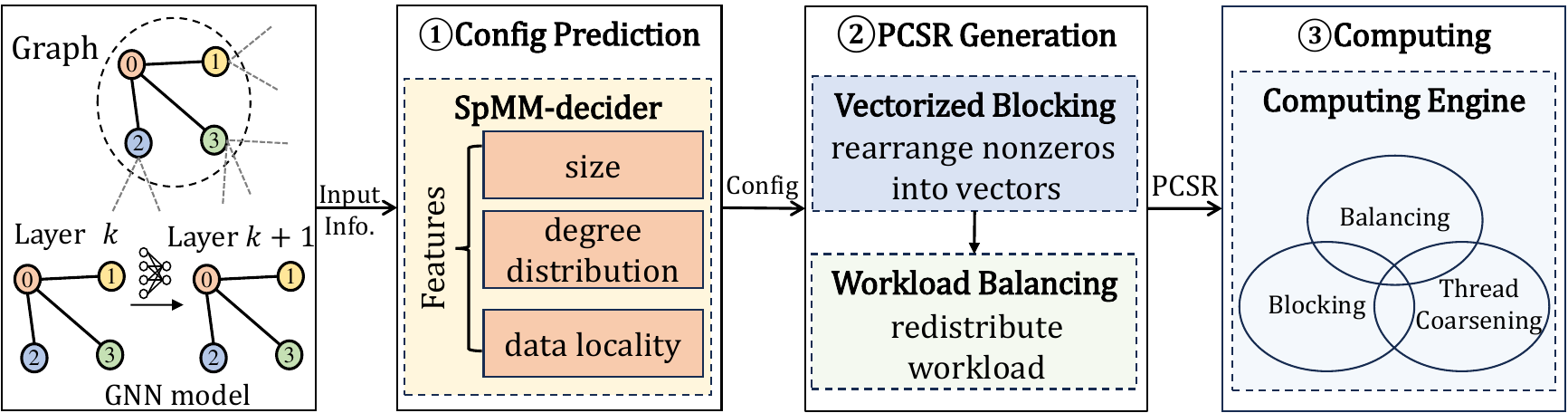}
    \vspace{-10pt}
    \caption{Overview of ParamSpMM: a three-phase workflow for SpMM in GNNs.}
    \label{fig:overview}
    \vspace{-10pt}
\end{figure*}

\vspace{-10pt}
\section{Backbone Design of ParamSpMM}\label{design:ParamSpMM}
\vspace{-10pt}
In this section, we first provide an overview of ParamSpMM. Then, we introduce Parameterized Compressed Sparse Row (PCSR), the data representation of a sparse matrix in ParamSpMM. Subsequently, we detail the computing engine and present graph reordering to further enhance ParamSpMM's performance.

\vspace{-10pt}
\subsection{Overview of ParamSpMM}\label{design:ParamSpMM-overview}
ParamSpMM consists of three phases: configuration prediction, PCSR generation, and SpMM computing (Figure \ref{fig:overview}). First, the ML-based SpMM-decider (Section \ref{design_SpMM_decider}) predicts optimal $\langle \mathcal{W}, \mathcal{F}, \mathcal{V}, \mathcal{S} \rangle$ using input features. Second, it applies vectorized blocking to extract nonzero vectors from $\bm{A}$, followed by workload balancing if $\mathcal{S}$ is $True$. Workload balancing redistributes workload across warps and generates output positions for partial result accumulation in $\bm{C}$. Finally, based on PCSR, the computing engine executes the SpMM kernel with customized optimization of blocking, workload balancing, and thread coarsening.

\begin{figure}[t]
    \centering
    \includegraphics[width=0.9\linewidth]{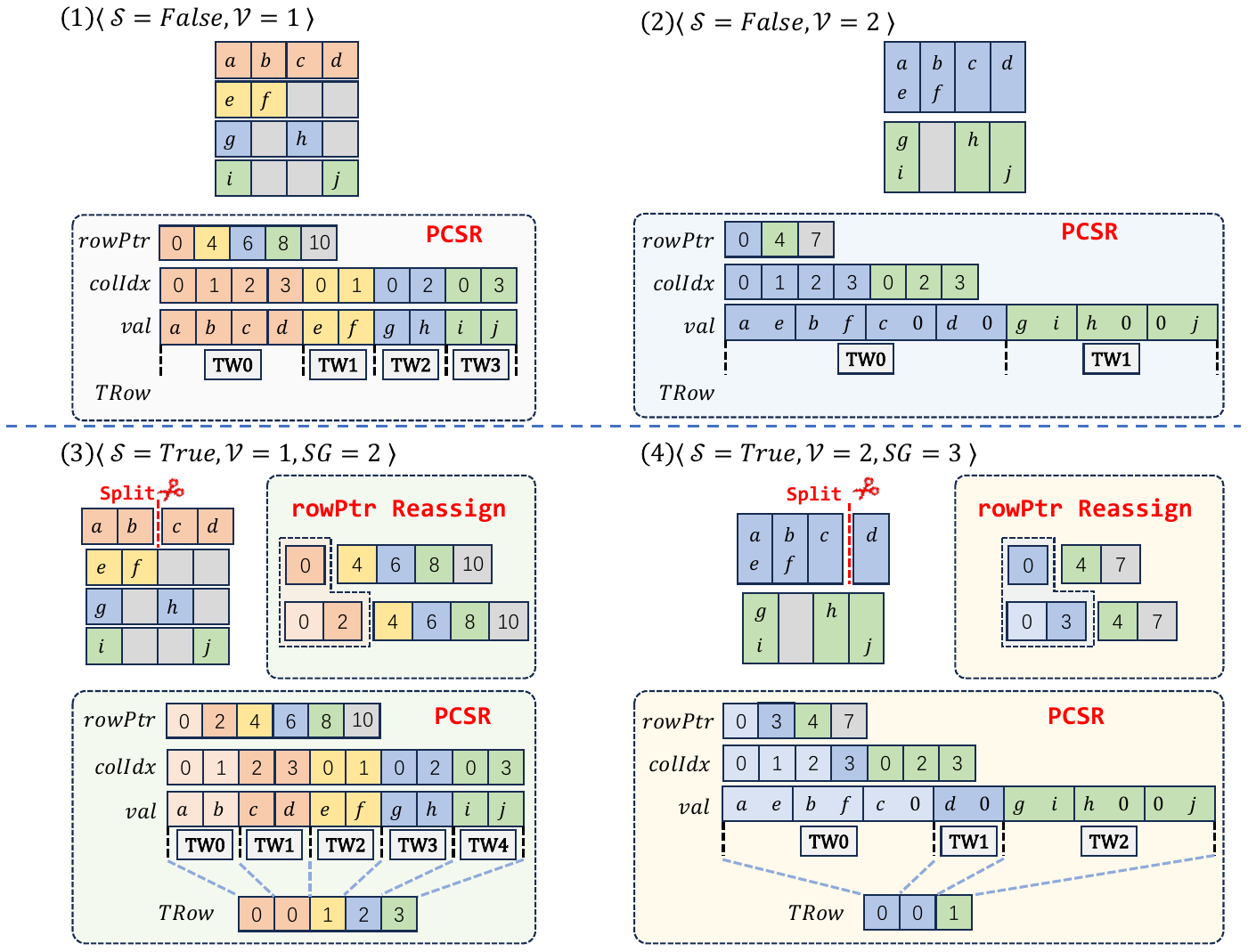}
    \vspace{-10pt}
    \caption{PCSR under four different configurations. TW: Thread Warp.}
    \label{fig:basic_config}
    \vspace{-15pt}
\end{figure}

\vspace{-10pt}
\subsection{Parameterized Compressed Sparse Row Format}
To enable seamless integration of various optimizations in ParamSpMM, we propose Parameterized Compressed Sparse Row (PCSR), an adaptive representation for sparse matrix $\bm{A}$.

\textbf{Data Representation of PCSR.} PCSR represents a sparse matrix via four arrays: $rowPtr$, $colIdx$, $val$, and $TRow$, arranging matrix elements into $\mathcal{V}\times1$ nonzero vectors. $rowPtr$ defines the traversal range of nonzero vectors and indicates each thread warp's workload $l$. $val$ and $colIdx$ store the values and column indices of nonzero vectors. $TRow$ is for workload balancing: without balancing, a thread warp determines its output position in $\bm{C}$ using $(blk.x,warpId)$ (Lines 3,11 of Algorithm~\ref{Alg:CSR_SpMM}), leaving $TRow$ empty. With balancing, the workload is redistributed across multiple warps, using $TRow$ to store target positions in $\bm{C}$ to accumulate partial results, and $rowPtr$ is reassigned to redistribute the workload across thread warps.

\textbf{PCSR Generation.} Given the configuration of $\langle \mathcal{W}, \mathcal{F}, \mathcal{V}, \mathcal{S} \rangle$, it begins by dividing $\bm{A}$ into multiple row panels of height $\mathcal{V}$, followed by vectorized blocking to generate $rowPtr$, $colIdx$, and $val$ for the nonzero vectors (size of $\mathcal{V}\times1$) within row panels. Subsequently, if $\mathcal{S}=True$, for workload balancing, $rowPtr$ is reassigned to redistribute the workload across thread warps. $TRow$ is generated to store the target positions in $\bm{C}$ for partial results accumulation.

\textbf{Examples of PCSR.} Figure \ref{fig:basic_config} illustrates the examples of PCSR under four different configurations, where different row panels are marked with different colors. In Examples $(2)$ and $(4)$, $val$ contains flattened vectors with some necessary zero padding when the number of nonzeros in a vector is less than $\mathcal{V}$. Elements such as $a$ and $e$ can thus reuse the data from dense matrix $\bm{B}$ when packed into a vector, attributed to their identical column index. Examples $(3)$ and $(4)$ with $\mathcal{S}=True$ are cases when workload balancing is applied. Example $(3)$ differs from Example $(1)$ by sharing the original workload of thread warp 0 with warp 1, thus improving workload balance. However, Example $(4)$ reveals a worse workload distribution after balancing, where the original workload in Example $(2)$ is already fairly balanced, highlighting the importance of proper $\mathcal{S}$ decisions as discussed in Section \ref{opp_balancing}.

\begin{algorithm}[h]
    \renewcommand{\algorithmicrequire}{\textbf{Input:}}
    \renewcommand{\algorithmicensure}{\textbf{Output:}}
    \caption{ParamSpMM Computing Engine}
    \label{Alg:ParamSpMM}
    \begin{algorithmic} [1]
        \REQUIRE{PCSR, $dim$, $\langle \mathcal{W}, \mathcal{F}, \mathcal{V}, \mathcal{S}  \rangle$}
        \ENSURE{$C$}
        \STATE \textcolor{blue}{/*initialization (Lines 1-6 of Algorithm~\ref{Alg:CSR_SpMM})*/}
        \STATE{$res[\mathcal{V}][\mathcal{F}]=\{0\}$}
        \STATE {$t = min(\mathcal{F},CEILDIV(dim -  seg, \omega$))}
        \FOR{ $i$ in $[head, tail)$}
        \STATE \textcolor{blue}{/*thread coarsening: prefetch $val$ and $Brow$*/}
        \STATE {$Brow =\ $PCSR.$colIdx[i]$}
        \STATE {$v[0:\mathcal{V})=\ $PCSR.$val[i*\mathcal{V}:(i+1)*\mathcal{V})$}
        \STATE \textcolor{blue}{/*performing $t$ MAC operations*/}
        \FOR{$j$ in $[0,t)$}
        \STATE \textcolor{blue}{/*blocking: prefetch $Bval$ for data reuse*/}
        \STATE{$Bval = \bm{B}[Brow][seg + j*\omega]$}
        \FOR{$k$ in $[0,\mathcal{V})$}
        \STATE{$res[k][j]+=v[k]*Bval$}
        \ENDFOR
        \ENDFOR
        \ENDFOR
        \IF{$\mathcal{S}==False$}
        \STATE{$Crow=Crow*\mathcal{V}$}
        \FOR{$i$ in $[0,t)$}
        \FOR{$j$ in $[0,\mathcal{V})$}
        \STATE{$\bm{C}[Crow+j][seg+i*\omega]=res[j][i]$}
        \ENDFOR
        \ENDFOR
        \ELSE
        \STATE{$Crow=\ $PCSR$.TRow[Crow]*\mathcal{V}$}
        \FOR{$i$ in $[0,t)$}
        \FOR{$j$ in $[0,\mathcal{V})$}
        \STATE{$\operatorname{atomicAdd}(res[j][i],\bm{C}[Crow+j][seg+i*\omega])$}
        \ENDFOR
        \ENDFOR
        \ENDIF
    \end{algorithmic}
\end{algorithm}

\vspace{-10pt}
\subsection{Computing Engine}\label{design:computing-engine}
\vspace{-5pt}
The computing engine executes SpMM using PCSR with a customized kernel based on the predicted $\langle \mathcal{W}, \mathcal{F}, \mathcal{V}, \mathcal{S} \rangle$ by SpMM-decider. Algorithm \ref{Alg:ParamSpMM} outlines the core functionality from the perspective of a thread in a warp.

The algorithm begins by defining the traversal ranges of nonzero vectors via PCSR$.rowPtr$, setting thread mapping with $\mathcal{W}$, and initializing $res$ buffer (Lines 1-2). The variable $t$ (Line 3) determines the actual MAC operations per thread in each nonzero vector iteration (Line 4). $t$ is crucial for handling the boundary cases when $dim$ is not a multiple of $\omega*\mathcal{F}$, since some threads may execute fewer than $\mathcal{F}$ MAC operations in one iteration.

Subsequently, it iterates through the allocated nonzero vectors (Line 4). For each vector, it performs thread coarsening by prefetching PCSR$.val$ and PCSR$.colIdx$ (Lines 6-7) for subsequent $t$ MAC operations within the nested loops (Lines 9-15). Through blocking, it reuses $\bm{B}$'s values $\mathcal{V}$ times (Lines 11-14), reducing memory transactions.

Lastly, if workload balancing is not applied ($\mathcal{S}=False$), the results are directly written to $\bm{C}$ (Lines 18-23). Otherwise, when $\mathcal{S}=True$, it uses the PCSR.$TRow$ to locate the write-back positions and performs $\operatorname{atomicAdd}$ to handle potential conflicts in partial results accumulation (Lines 25-30).

\vspace{-10pt}
\subsection{Graph Reordering to Enhance Data Locality}
Various graph reordering techniques \cite{arai2016rabbit, huang2021understanding, wang2021gnnadvisor, fan2023fast, reorder_wei} have been devised to improve data locality by rearranging nodes with similar neighbors, which share similar memory access patterns for $\bm{B}$ during computation, to be positioned closer. This complements ParamSpMM of $\mathcal{V}>1$ by creating more consecutive nonzeros in the same columns, leading to less zero padding. Appropriate application of blocking can further exploit the enhanced data locality from graph reordering. Practically speaking, Rabbit Reordering \cite{arai2016rabbit} is a default step in ParamSpMM, which is highly parallelized and time-efficient. The reordering cost is amortizable as reordered graphs can be reused in iterative applications.

\vspace{-10pt}
\section{Learning Optimal ParamSpMM Configuration}\label{design_SpMM_decider}
As the optimal ParamSpMM configuration varies across SpMM inputs and is challenging to determine manually, we propose a data-driven ML-based SpMM-decider to predict the optimal configuration. In this section, we introduce our carefully crafted sparse matrix features and the configuration prediction model.

\vspace{-10pt}
\subsection{Sparse Matrix Features}
We classify sparse matrix features into three categories: (1) Size features: They decide the scale of threads and overall workload, guiding the setting of $\mathcal{F}$ and $\mathcal{W}$; (2) Degree distribution features: They navigate the decision of workload balancing; (3) Data locality features: They help the selection of $\mathcal{V}$ in vectorized blocking. These features, detailed in Table \ref{tab:Summary_Features}, can be measured once to train models and configure ParamSpMM across different $dim$, which is amortizable in iterative applications. While most features are well-known, we specifically introduce two essential metrics, $SR$ and $PR$, to guide ParamSpMM's optimization of blocking and balancing:

\vspace{-10pt}
\subsubsection{$PR$ for Blocking.}
As discussed in Section \ref{opp_blocking}, $\mathcal{V}$ in vectorized blocking significantly impacts zero padding levels, consequently affecting ParamSpMM performance. To quantify the zero padding level, we introduce $PR_{\mathcal{V}}$:
\begin{equation}
    \vspace{-8pt}
    PR_{\mathcal{V}}=1- \frac{nnz}{nnz_{\mathcal{V}}* \mathcal{V}}\label{eq:reuseRate}
\end{equation}
where $nnz$ is the number of nonzeros and $nnz_{\mathcal{V}}$ is the number of nonzero vectors. $PR_{\mathcal{V}}$ ranges from $[0,1-\frac{1}{\mathcal{V}}]$, with higher values indicating increased zero padding. Lower $PR_{\mathcal{V}}$ values reflect better data reuse and reduced unnecessary computation, potentially leading to improved performance. Consequently, $PR_{\mathcal{V}}$ serves as one of many input features for the ML-based SpMM-decider.

\vspace{-10pt}
\subsubsection{$SR$ for Balancing.}
Our workload balancing method sets a workload upper bound, Split Granularity $(SG)$, ensuring each warp processes no more than $SG$ nonzero vectors. We set $SG$ as:
\begin{equation}\label{eq_SG}
    \vspace{-8pt}
    SG= CEILDIV(\widehat{d_V},\omega)*\omega
\end{equation}
where $\omega$ is the thread warp size. After balancing, multiple warps may accumulate partial results to the same segment in $\bm{C}$, increasing memory writes as discussed in Section \ref{opp_balancing}. To quantify this overhead, we introduce Split Ratio ($SR$):
\begin{equation}
    \vspace{-8pt}
    SR=\frac{split\_size}{row\_size}\label{eq_SR}
\end{equation}
where $split\_size$ and $row\_size$ are the lengths of reassigned and original $rowPtr$ arrays, respectively. $SR \in [1,+\infty]$ measures the increased memory writes to $\bm{C}$. For example, $SR=1.2$ indicates $\mathbf{1.2\times}$ more writes than without balancing. Given the complexity of manually selecting the optimal strategy based on $SR$, the ML-based SpMM-decider is employed for this decision.

\begin{table}[t]
    \centering
    \vspace{-10pt}
    \caption{Summary of Features}
    \vspace{-10pt}
    \label{tab:Summary_Features}
    \begin{tabular}{c|ll}
        \toprule
        \multicolumn{1}{l|}{\textbf{Category}}               & \textbf{Feature} & \textbf{Description}                                                                                                                                                                                             \\ \midrule
        \multirow{7}{*}{\textbf{\begin{tabular}[c]{@{}c@{}}size\\ features\end{tabular}}} & $n$              & number of rows in a sparse matrix                                                                                                                                                                                \\
                                                             & $\widehat{n}$    & number of non-empty rows in a sparse matrix                                                                                                                                                                      \\
                                                             & $nnz$            & number of nonzeros in a sparse matrix                                                                                                                                                                            \\
                                                             & $\delta $        & the ratio of $\widehat{n}$ to $n$                                                                                                                                                                                \\
                                                             & $d$              & the average number of nonzeros per row                                                                                                                                                                           \\
                                                             & $\widehat{d}$    & $d$ without considering empty rows                                                                                                                                                                               \\
                                                             & $d_{max}$        & max number of nonzeros across sparse rows                                                                                                                                                                        \\ \midrule
        \multirow{3}{*}{\textbf{\begin{tabular}[c]{@{}c@{}}degree \\ distribution \\ features\end{tabular}}} & $CV$             & coefficient of variation of node degrees                                                                                                                                                                         \\
                                                             & $\widehat{CV}$   & $CV$ without considering empty rows                                                                                                                                                                              \\
                                                             & $SR_i$           & $SR$  under $\langle \mathcal{V}=i,\mathcal{S}= true \rangle$                                                                                                                                                    \\ \midrule
        \multirow{4}{*}{\textbf{\begin{tabular}[c]{@{}c@{}}data\\ locality\\ features\end{tabular}}} & $\rho$           & the density of nonzeros in a sparse matrix                                                                                                                                                                       \\
                                                             & $b$              & average bandwidth of all rows                                                                                                                                                                                    \\
                                                             & $b_{max}$        & max bandwidth\tablefootnote{ We define the bandwidth of one row in the sparse matrix as the difference in column indices between the first nonzero element and the last one.} across all rows in a sparse matrix \\
                                                             & $PR_i$           & padding ratio under vectorized blocking of $\mathcal{V}=i$                                                                                                                                                       \\ \bottomrule
    \end{tabular}%
    \vspace{-15pt}
\end{table}

\vspace{-12pt}
\subsection{SpMM-decider: Configuration Prediction Model}\label{decider:features&ML}
\vspace{-5pt}
Despite different ParamSpMM configurations serving distinct optimization purposes, establishing a definitive boundary among their application contexts is challenging. An ML-based SpMM-decider is thus trained to predict optimal ParamSpMM configurations based on the crafted input features, enabling systematic and data-driven SpMM optimization.

With the sparse matrix features listed in Table \ref{tab:Summary_Features} as input, ML-based SpMM-decider predicts the optimal configuration of $\langle \mathcal{W}, \mathcal{F}, \mathcal{V}, \mathcal{S}  \rangle$. SpMM-decider is based on the random forests model, which is a lightweight ensemble learning model. This approach facilitates easier model training and deployment, as well as a lower risk of overfitting compared to other more complicated ML models.

\begin{figure*}[t]
    \centerline{\includegraphics[width=0.99\linewidth]{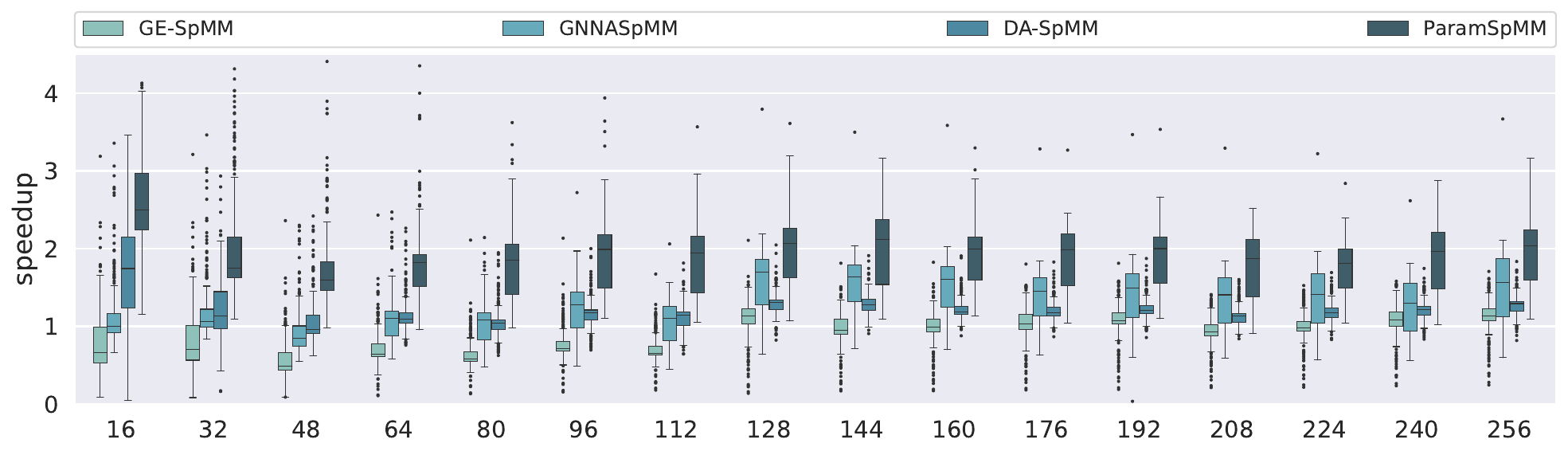}}
    \vspace{-10pt}
    \caption{The performance of SpMM across various $dim$. The reported speedup is normalized to \textbf{cuSPARSE}.}
    \label{fig:SpMM_bench}
    \vspace{-15pt}
\end{figure*}

\vspace{-5pt}
\section{Experimental Evaluation}
In this section, we evaluate ParamSpMM’s performance, technique effectiveness, and application in GNN training.

\vspace{-10pt}
\subsection{Experimental Setup} \label{exp_setup}
\paragraph{Experiment Platform.} We evaluate ParamSpMM and baselines on NVIDIA A6000, which has 84 Amper SMs and a compute capability of 8.6. We integrate ParamSpMM as an extension into PyTorch, leveraging its expressiveness in GNNs. The code is available at \url{https://github.com/zzzlxhhh/ParamSpMM_AE}, which is compiled with GCC 9.4.0 and NVCC 11.6 using -O3 optimization.

\vspace{-5pt}
\paragraph{Datasets.} For SpMM datasets, we collect 202 matrices ($1,000<n<8,000,000$) from SNAP \cite{leskovec2014snap} and DIMACS10 \cite{bader201110thDIMACS}. Too small matrices are exempted for the evaluation consistency concerns. Matrices that are too large are excluded due to possible out-of-memory errors. For GNN datasets, we take 6 graphs from OGB \cite{hu2020OGB}, which have not previously been used for training SpMM-decider.

\vspace{-5pt}
\paragraph{SpMM Baselines.} We compare ParamSpMM with three categories of baselines: \textbf{Static} (Nvidia cuSPARSE \cite{naumov2010cusparse} and GE-SpMM \cite{huang2020ge}), \textbf{Heuristic-based} (GNNAdvisor \cite{wang2021gnnadvisor}), and \textbf{ML-based} (DA-SpMM \cite{dai2022heuristic-DASpMM}). All methods use Rabbit Reordering \cite{arai2016rabbit} as default preprocessing.

\vspace{-5pt}
\paragraph{GNN Evaluation Models.} We evaluate ParamSpMM on two representative GNN models: Graph Convolutional Network (GCN)~\cite{GCN} and Graph Isomorphic Network (GIN)~\cite{GIN}. We use GNNAdivisor \cite{wang2021gnnadvisor} (a GNN system with optimized SpMM kernel) and DGL \cite{DGL} (a vendor-provided GNN library) as baselines.

\vspace{-5pt}
\subsection{Overall ParamSpMM Performance}\label{exp_overall}
We evaluate ParamSpMM's performance across 202 matrices with varying $dim$, resulting in 3232 diverse SpMM inputs exhibiting a substantial diversity. These matrices exhibit diverse characteristics, including matrix size ($n$ from $1005 \sim 7,733,822$), density ($\rho$ from $2.73\times e^{-7} \sim 0.025$), coefficient of variation ($CV$ from $0.00638 \sim 58.097$), zero padding ratio ($PR_2$ from $0.247 \sim 0.499$).

To visualize the adaptability of ParamSpMM and the baselines across various SpMM inputs, Figure \ref{fig:SpMM_bench} presents the boxplot of the speedups of each method over cuSPARSE on A6000, demonstrating the speedups distribution of each method. ParamSpMM significantly outperforms all baselines across a wide range of SpMM input diversity, demonstrating superior adaptability. While baseline methods occasionally underperform cuSPARSE, ParamSpMM maintains robust acceleration in most cases. Table~\ref{tab_avg_speedup} further summarizes the speedups of ParamSpMM over baselines across various $dim$. ParamSpMM achieves an average speedup of $\mathbf{1.92\times}$, $\mathbf{2.41\times}$, $\mathbf{1.55\times}$, and $\mathbf{1.64\times}$ over cuSPARSE, GE-SpMM, GNNAdivisor, and DA-SpMM, demonstrating superior performance.

\begin{table*}[h]
    \small
    \vspace{-10pt}
    \centering
    \caption{The speedups of ParamSpMM over baselines on A6000.}
    \label{tab_avg_speedup}
    \vspace{-5pt}
    \resizebox{1.0\textwidth}{!}{
        \begin{tabular}{c|c|cccccccccccccccc|c}
            \toprule
            \textbf{Category}                           & \textbf{Baselines}   & \textbf{16}  & \textbf{32}  & \textbf{48}  & \textbf{64}  & \textbf{80}  & \textbf{96}  & \textbf{112} & \textbf{128} & \textbf{144} & \textbf{160} & \textbf{176} & \textbf{192} & \textbf{208} & \textbf{224} & \textbf{240} & \textbf{256} & \textbf{Average}  \\ \midrule
            \multirow{2}{*}{\centering \textbf{Static}} & \textbf{cuSPARSE}    & $2.68\times$ & $2.07\times$ & $1.73\times$ & $1.81\times$ & $1.77\times$ & $1.90\times$ & $1.84\times$ & $1.97\times$ & $2.00\times$ & $1.90\times$ & $1.88\times$ & $1.87\times$ & $1.77\times$ & $1.75\times$ & $1.87\times$ & $1.93\times$ & $\bm{1.92\times}$ \\
                                                        & \textbf{GE-SpMM}     & $4.32\times$ & $3.03\times$ & $3.30\times$ & $2.79\times$ & $3.00\times$ & $2.67\times$ & $2.78\times$ & $2.00\times$ & $2.24\times$ & $2.08\times$ & $1.96\times$ & $1.88\times$ & $1.98\times$ & $1.88\times$ & $1.84\times$ & $1.82\times$ & $\bm{2.41\times}$ \\ \midrule
            \textbf{Heuristic}                          & \textbf{GNNAdivisor} & $2.44\times$ & $1.70\times$ & $1.87\times$ & $1.67\times$ & $1.71\times$ & $1.57\times$ & $1.81\times$ & $1.29\times$ & $1.31\times$ & $1.28\times$ & $1.39\times$ & $1.54\times$ & $1.34\times$ & $1.33\times$ & $1.54\times$ & $1.34\times$ & $\bm{1.55\times}$ \\ \midrule
            \textbf{ML}                                 & \textbf{DA-SpMM}     & $2.10\times$ & $1.95\times$ & $1.64\times$ & $1.58\times$ & $1.69\times$ & $1.64\times$ & $1.66\times$ & $1.53\times$ & $1.57\times$ & $1.58\times$ & $1.58\times$ & $1.54\times$ & $1.57\times$ & $1.49\times$ & $1.56\times$ & $1.52\times$ & $\bm{1.64\times}$ \\ \bottomrule
        \end{tabular}
    }
    \vspace{-10pt}
\end{table*}

\vspace{-20pt}
\subsection{Effectiveness of SpMM-decider}
\vspace{-5pt}
To evaluate SpMM-decider's prediction accuracy, we split the datasets into $\mathbf{80\%}$ training and $\mathbf{20\%}$ testing sets. We compare the performance of predicted configurations (\texttt{pre}) against optimal configurations, with randomly configured ParamSpMM (\texttt{rnd}) serving as a baseline. Table \ref{tab_norm_perf} shows that SpMM-decider effectively provides suitable configurations, with most normalized performances exceeding $\mathbf{99\%}$, whereas \texttt{rnd} shows significantly lower performance.

\begin{table}[h]
    \vspace{-10pt}
    \centering
    \caption{The normalized performance of SpMM-decider}
    \label{tab_norm_perf}
    \vspace{-5pt}
    \resizebox{0.75\columnwidth}{!}{%
        \begin{tabular}{c|cc|c|cc|c|cc|c|cc}
            \toprule
            \textbf{dim} & \texttt{pre} & \texttt{rnd} & \textbf{dim} & \texttt{pre} & \texttt{rnd} & \textbf{dim} & \texttt{pre} & \texttt{rnd} & \textbf{dim} & \texttt{pre} & \texttt{rnd} \\ \midrule
            16           & 98.84\%      & 82.26\%      & 80           & 99.97\%      & 70.58\%      & 144          & 99.28\%      & 74.04\%      & 208          & 99.29\%      & 68.60\%      \\
            32           & 99.69\%      & 73.47\%      & 96           & 99.55\%      & 75.94\%      & 160          & 99.74\%      & 72.13\%      & 224          & 99.67\%      & 79.42\%      \\
            48           & 99.98\%      & 70.34\%      & 112          & 99.31\%      & 70.45\%      & 176          & 99.21\%      & 73.49\%      & 240          & 99.19\%      & 70.56\%      \\
            64           & 98.24\%      & 76.65\%      & 128          & 99.30\%      & 78.14\%      & 192          & 98.96\%      & 74.18\%      & 256          & 98.75\%      & 70.33\%      \\
            \bottomrule
        \end{tabular}%
    }
    \vspace{-10pt}
\end{table}

\vspace{-15pt}
\subsection{Effectiveness of Graph Reordering}
\vspace{-5pt}
In this section, we evaluate the impact of graph reordering. With ParamSpMM\_wor and cuSPARSE\_wor to denote ParamSpMM and cuSPARSE without reordering, Table~\ref{tab_reorder} reveals three key findings: (1) Graph reordering significantly enhances ParamSpMM's performance over ParamSpMM\_wor. (2) Even without reordering, ParamSpMM\_wor still substantially outperforms cuSPARSE\_wor due to our parametric approach. (3) While graph reordering provides cuSPARSE a modest $\bm{1.14\times}$ speedup over cuSPARSE\_wor, ParamSpMM better leverages the improved data locality, achieving a $\bm{1.26\times}$ speedup over ParamSpMM\_wor.

\begin{table}[h]
    \small
    \centering
    \vspace{-10pt}
    \caption{The speedups of cuSPARSE, ParamSpMM, and ParamSpMM\_wor over cuSPARSE\_wor.}
    \vspace{-5pt}
    \label{tab_reorder}
    \resizebox{0.75\columnwidth}{!}{%
        \begin{tabular}{c|cccccccc|c}
            \toprule
            \textbf{dim}            & \textbf{16}  & \textbf{32}  & \textbf{48}  & \textbf{64}  & \textbf{80}  & \textbf{96}  & \textbf{112} & \textbf{128} & \textbf{Average}  \\ \midrule
            \textbf{cuSPARSE}       & $1.06\times$ & $1.08\times$ & $1.15\times$ & $1.16\times$ & $1.18\times$ & $1.17\times$ & $1.17\times$ & $1.15\times$ & $\bm{1.14\times}$ \\
            \textbf{ParamSpMM\_wor} & $1.91\times$ & $1.92\times$ & $1.66\times$ & $1.70\times$ & $1.66\times$ & $1.72\times$ & $1.68\times$ & $1.76\times$ & $\bm{1.75\times}$ \\
            \textbf{ParamSpMM}      & $2.80\times$ & $2.19\times$ & $1.95\times$ & $2.07\times$ & $2.06\times$ & $2.20\times$ & $2.13\times$ & $2.26\times$ & $\bm{2.21\times}$ \\
            \bottomrule
        \end{tabular}
    }
    \vspace{-15pt}
\end{table}

\vspace{-15pt}
\subsection{Application in GNNs}\label{exp:GCN}
\begin{figure}[h]
    \vspace{-15pt}
    \centering
    \includegraphics[width=0.9\linewidth]{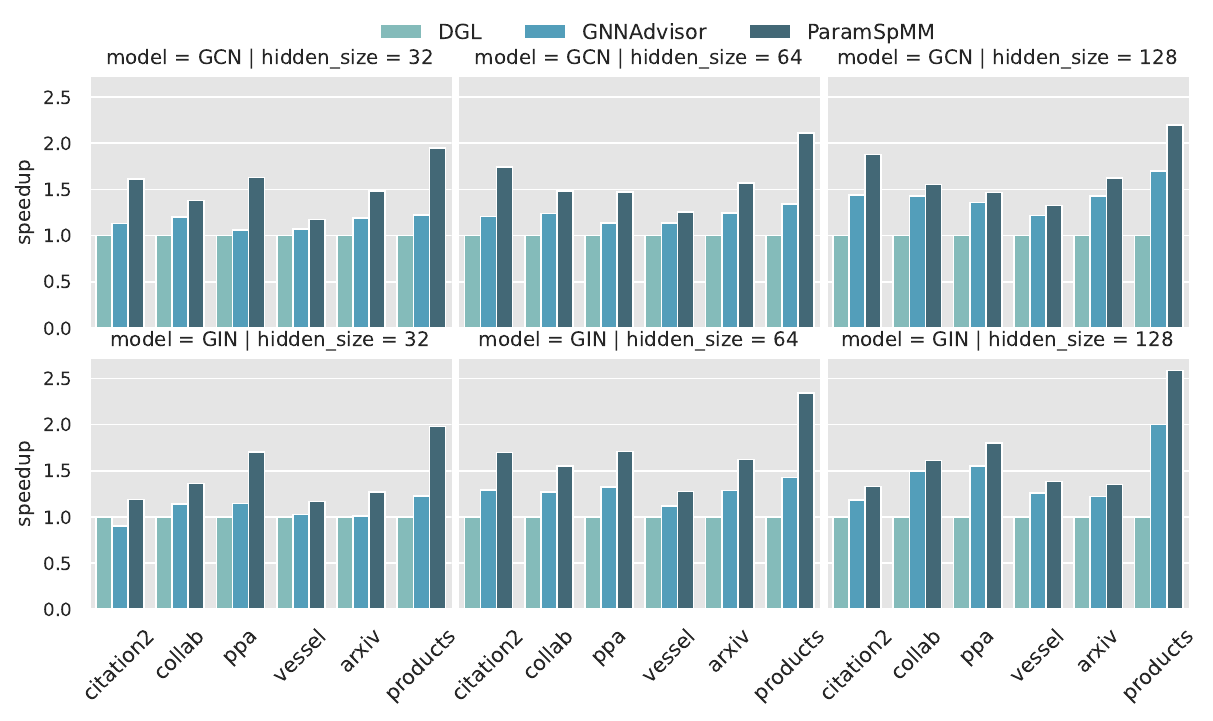}
    \vspace{-10pt}
    \caption{GCN and GIN performance speedup normalized to DGL.}
    \label{fig_GCN_GIN}
    \vspace{-10pt}
\end{figure}

We evaluate ParamSpMM in 5-layer GCN and GIN with input/output sizes of 16 and hidden sizes of $\{32,64,128\}$ on A6000. Figure \ref{fig_GCN_GIN} shows training speedups of ParamSpMM and GNNAdvisor over DGL. ParamSpMM consistently outperforms DGL, achieving average speedups of $\mathbf{1.60\times}$ (up to $\mathbf{2.19\times}$) for GCN and $\mathbf{1.61\times}$ (up to $\mathbf{2.59\times}$) for GIN. As SpMM is key in both models, ParamSpMM effectively accelerates both GCN and GIN.

\vspace{-10pt}
\section{Related Work}
\vspace{-10pt}
In this section, we discuss the related works for SpMM optimization.

\textbf{Blocking.} Blocking techniques~\cite{li2022efficient-magicube, ASpT:hong2019adaptive,yesil2022-DDB,flash-LLM} exploit data locality. FlashLLM \cite{flash-LLM} and Magicube~\cite{li2022efficient-magicube} utilize the high throughput tensor core units to facilitate SpMM in sparse DNN training~\cite{hoefler2021sparsityDNN}, where the sparse matrices are relatively dense and uniform. Other studies~\cite{yesil2022-DDB, ASpT:hong2019adaptive,vldb-SpMV} propose hybrid blocking strategies with better adaptability to various data locality.

\textbf{Workload Balancing.} While merge-based balancing~\cite{yang2018design_principles,dai2022heuristic-DASpMM,merrill2016mergeSpMV} accelerates SpMM for Power-law distributions, it requires costly binary searches for nonzeros' row indices. GNNAdvisor~\cite{wang2021gnnadvisor} proposes a complex shared memory accumulation mechanism to reduce global memory atomic operations after balancing.

\textbf{Thread Coarsening.} While several works~\cite{wang2021gnnadvisor, huang2020ge, fan2023fast} employ thread coarsening to reduce memory access, they fail to configure an appropriate $\mathcal{F}$. GNNAdvisor~\cite{wang2021gnnadvisor} and GE-SpMM~\cite{huang2020ge} simply increase $\mathcal{F}$ with $dim$ without considering MAC-job gaps. HP-SpMM~\cite{fan2023fast} applies vectorized instructions (float2 and float4) to achieve thread coarsening, leading to limited $\mathcal{F}$ choices from 2 and 4.

\textbf{ML-guided Optimization.} ML is used as an auto-tuning approach in some recent works~\cite{yesil2022-DDB, dai2022heuristic-DASpMM}. DA-SpMM~\cite{dai2022heuristic-DASpMM} utilizes ML to optimize SpMM on GPUs, yet their strategy space overlooks blocking and thread coarsening. DDB~\cite{yesil2022-DDB} focuses on hardware with matrix-multiply units.

\vspace{-10pt}
\section{Conclusions}
\vspace{-5pt}
In this work, we presented ParamSpMM, a flexible parametric framework for optimizing GPU-based SpMM kernels against diverse inputs in GNNs. After analyzing existing works' limitations in handling input diversity, we flexibly integrated blocking, workload balancing, and thread coarsening techniques, offering customizable optimization through parameter configuration. We introduced PCSR format to enable seamless cooperation among optimization techniques. We developed an ML-based SpMM-decider for automatic configuration prediction based on input characteristics. Extensive evaluations show ParamSpMM's superior adaptability with an average $\bm{1.92\times}$ speedup over cuSPARSE~\cite{naumov2010cusparse} while achieving significant acceleration over DGL~\cite{DGL} in various GNN models.

\vspace{-10pt}
\section{Acknowledgments}
\vspace{-5pt}
This work is supported by National Natural Science Foundation of China (Nos. 62272054, 62192784, 62372055), Beijing Nova Program (No. 20230484319, 20250484968), State Key Laboratory of Multimedia Information Processing Open Fund (No. SKLMIP-KF-2025-07),  and CAAI-CANN Open Fund, developed on OpenI Community (No. CAAIXSJLJJ2025CANN10). Yingxia Shao is the corresponding author.

%
%
%
\vspace{-10pt}
\bibliographystyle{splncs04}
\bibliography{base}

\end{document}